\documentclass[12pt]{article}
\usepackage{authblk}
\usepackage{hyperref}
\hypersetup{
    colorlinks=true,
    linkcolor=blue,     
    urlcolor=cyan,%cyan,
    citecolor=green
    }
\usepackage[top=25truemm,bottom=28truemm,left=25truemm,right=25truemm]{geometry}
\usepackage{setspace}
\usepackage{amsmath,amssymb,mathrsfs,amsbsy,latexsym,amsfonts,amsthm}
\usepackage{fancyhdr}
\pdfoutput=1 % For pdfLaTeX
\usepackage{graphicx}
\usepackage{xcolor}
\usepackage{comment}
\usepackage{here}
\usepackage{braket}
\usepackage[utf8]{inputenc}
\usepackage{tensor}
\usepackage{mathtools}

\newcommand{\nn}{\nonumber\\}

\newcommand{\tk}{\tilde{k}}

\newcommand{\cO}{\mathcal{O}}
\newcommand{\cN}{\mathcal{N}}
\newcommand{\op}{\omega_p}
\newcommand{\ok}{\omega_k}

\numberwithin{equation}{section}
\title{\bf Notes on Rindler wave packets\\
in Minkowski spacetime}

\author[1]{ 
	Shono~Shibuya\thanks{\tt shibuya.shono.n8(at)s.mail.nagoya-u.ac.jp}
}

\author[2, 3]{
Sotaro~Sugishita\thanks{\tt sotaro(at)gauge.scphys.kyoto-u.ac.jp}
}

\vspace{5mm}
\affil[1]{\it\normalsize Department of Physics, Nagoya University, Nagoya, Aichi 464-8602, Japan}
\affil[2]{\it\normalsize Department of Physics, Kyoto University, Kyoto 606-8502, Japan}
\affil[3]{\it\normalsize RIKEN iTHEMS, Wako, Saitama 351-0198}
\date{}
\begin{document}

\maketitle
\thispagestyle{fancy}
\renewcommand{\headrulewidth}{0pt}

%%%%%%%%%%%%%%%%%%%%%%%%%%%%%%%%%%%%%%%%%%%%%%%%%%%%%%%%%%%%%%
%%%%%%%%%%%%%%%%%%%%%%%%%%%%%%%%%%%%%%%%%%%%%%%%%%%%%%%%%%%%%%
\begin{abstract}
We consider wave packets of a massless scalar field that have well-localized Rindler energy, and examine how their energy appears to a Minkowski observer to study how the classical gravitational red-shift formula is modified quantum mechanically.
We derive, by using the saddle point approximation, an analytic expression for the Minkowski momentum distribution of such Rindler wave packets.
We find a universal lower bound on the uncertainty in the Minkowski momentum; the momentum distribution can never become arbitrarily sharp.
\end{abstract}
%%%%%%%%%%%%%%%%%%%%%%%%%%%%%%%%%%%%%%%%%%%%%%%%%%%%%%%%%%%%%%
%%%%%%%%%%%%%%%%%%%%%%%%%%%%%%%%%%%%%%%%%%%%%%%%%%%%%%%%%%%%%%

\newpage
\thispagestyle{empty}
\setcounter{tocdepth}{2}

\setlength{\abovedisplayskip}{12pt}
\setlength{\belowdisplayskip}{12pt}

\tableofcontents
\newpage
%%%%%%%%%%%%%%%%%%%%%%%%%%%%%%%%%%%%%%%%%%%%%%%%%%%%%%%%%%%%%%
%%%%%%%%%%%%%%%%%%%%%%%%%%%%%%%%%%%%%%%%%%%%%%%%%%%%%%%%%%%%%%
\section{Introduction}
\label{sec:intro}
%%%%%%%%%%%%%%%%%%%%%%%%%%%%%%%%%%%%%%%%%%%%%%%%%%%%%%%%%%%%%%
%%%%%%%%%%%%%%%%%%%%%%%%%%%%%%%%%%%%%%%%%%%%%%%%%%%%%%%%%%%%%%

Gravitational redshift is a characteristic phenomenon of classical general relativity. 
Because time flows differently at different points in curved spacetime, the frequency of light depends on position. 
Even in flat spacetime, clocks of inertial and uniformly accelerated observers run at different rates, producing a red/blue-shift. 
This is a classical result, and it is natural to ask how it changes in quantum theory.

Hence, we aim to investigate the red/blue-shift effects in quantum field theories (QFTs). 
Classically, the observer-dependent redshift of a particle momentum can be evaluated by a simple coordinate transformation once the particle’s position is fixed. 
Quantum mechanically, however, there is no state in which both position and momentum are simultaneously specified with arbitrary precision; for instance, a momentum eigenstate (a plane wave) is completely delocalized in space. 
One must work with wave packets localized well in both variables to discuss the observer-dependent red/blue-shift quantum mechanically.
As a first step, we consider a wave packet state localized both in position and momentum for accelerating observers (Rindler observers), and examine how it appears to a Minkowski observer.

One motivation for this setup is that we are interested in the red/blue-shift caused by the horizon.
Near the horizon of a black hole, energy is strongly blue-shifted. 
A signal from a near-horizon region can appear low-energy to a distant observer, though it carries very high energy in the local proper frame near the horizon. 
Close enough to the horizon, the energy can exceed the Planck scale, where an ordinary QFT can no longer be trusted. 
This tension between low energies for a distant observer and trans-Planckian energies for a near-horizon observer is known as the trans-Planckian problem \cite{tHooft:1984kcu, Jacobson:1991gr}, and this problem has been discussed in the context of the Hawking radiation \cite{Hawking:1974rv, Hawking:1975vcx}. 
A similar trans-Planckian problem also arises in cosmology \cite{Jacobson:1999zk, Martin:2000xs}.

Although there are numerous studies claiming that the trans-Planckian problem of the Hawking radiation is not a genuine issue (see, e.g., \cite{Jacobson:1993hn, Unruh:1994je, Brout:1995wp, Hambli:1995pp, Himemoto:1999kd, Jacobson:1999zk, Unruh:2004zk}), recent active research \cite{Ho:2021sbi, Ho:2022gpg, Akhmedov:2023gqf, Chau:2023zxb, Ho:2023tdq, Ho:2024tby} suggests that the Hawking radiation may cease well before the Page time due to UV effects (see also \cite{Blamart:2023ixr}).
A crucial ingredient in these recent arguments is the use of localized wave packets \cite{Ho:2021sbi, Ho:2022gpg, Akhmedov:2023gqf, Chau:2023zxb, Ho:2023tdq, Ho:2024tby}.
In the earlier discussions of the trans-Planckian problem, wave-packet effects appear to have been overlooked, except for recent studies, e.g., \cite{Terashima:2020uqu, Terashima:2021klf, Ho:2021sbi, Sugishita:2022ldv, Ho:2022gpg, Sugishita:2023wjm}, 
although the importance of using wave packets in quantum field theory on curved spacetime has long been recognized (for instance, \cite{Hawking:1975vcx}).\footnote{
The importance of wave packets has also been pointed out recently in, e.g., \cite{Ishikawa:2018koj, Hanada:2021ipb, Ishikawa:2021bzf, Oda:2021tiv, Edery:2021thf, Gautam:2022akq, Oda:2023qek, Ishikawa:2023bnx, Gautam:2024zsj}.
Wave packets in AdS space have also been investigated (see, e.g., \cite{Terashima:2021klf, Sugishita:2022ldv, Kinoshita:2023hgc, Terashima:2023mcr, Tanahashi:2025fqi})}
Thus, despite its clear significance, the detailed analysis of wave packets on curved spacetime is still relatively unexplored.
This is one of the reasons why we examine the red/blue-shift due to the horizon in quantum theory using wave packets in this paper. 

In this paper, we ask a simple question: we revisit the classical red/blue-shift formula within the framework of QFT by tracking, via localized wave packets, how the energy seen by a Rindler observer appears to a Minkowski observer, and determine when the classical red/blue-shift formula is reproduced. 
To the best of our knowledge, this simple question has not been addressed previously.

Our result is related to the Unruh effect \cite{Fulling:1972md, Davies:1974th, Unruh:1976db}.
The statement that the Minkowski vacuum behaves as a thermal state for Rindler observers is likely true formally for any well-behaved UV-complete quantum field theories (without gravity) \cite{Bisognano:1976za, Unruh:1983ac}.
However, if we adopt the view that QFT is an effective field theory (EFT) that applies below some cutoff, it is unclear whether the statement is true. 
In the Unruh effect, the mixing of the modes by the Bogoliubov transformations plays the key role, and the transformation mixes the low-energy modes for an Rindler observer with the high-energy modes for the Minkowski observer. Thus, the thermal property of the Minkowski vacuum for the Rindler observer might get corrections if the theory has a cutoff.
In the real world, gravity exists, and any QFT description without gravity must have a cutoff near the Planck scale (or another scale).
At the Planck scale, even the idea of a smooth spacetime may fail. 
We have no guarantee that ordinary QFT remains valid near the Planck scale.\footnote{This trans-Planckian problem also causes issues in the treatment of the bulk subregion in the AdS/CFT correspondence, e.g., the AdS-Rindler reconstruction (or the entanglement wedge reconstruction) as claimed in \cite{Terashima:2020uqu, Terashima:2021klf, Sugishita:2022ldv, Sugishita:2023wjm, Sugishita:2024lee}.}
From this viewpoint, knowing how the classical redshift formula between Rindler and Minkowski changes in QFT is also crucial for identifying possible limits of the EFT description for the Unruh effect.

Here we briefly summarize our results. For simplicity, we consider a free massless scalar in two-dimensional spacetime.
We study how a wave packet with average energy $\omega_R$ measured in the Rindler time (which corresponds to the proper time for a distant observer of a black hole), appears to a Minkowski-time observer. 
We thus assume $\omega_R \gg a$ and $\omega_R \gg \Delta \omega_R$, where $a$ is the acceleration parameter of the Rindler coordinates and sets the scale of the Unruh temperature, and $\Delta \omega_R$ is the width of the Rindler energy for the wave packet. 

Under these conditions, we confirm that the classical blue-shift formula holds if $\Delta \omega_R \gg a$. 
However, when $\Delta \omega_R$ is reduced to $\mathcal{O}(a)$, the blue-shift gets enhanced compared to the classical result, worsening the trans-Planckian problem. 
Thus, to study modes with localized momenta $\Delta \omega_R \sim 0$ like plane waves, the EFT description is no longer valid, and we need a theory of quantum gravity.
In addition, we find that for any $\Delta \omega_R$, the corresponding width of the Minkowski energy $\Delta \omega_M$ has a lower bound and cannot be made arbitrarily small.
This is due to the fact that the time-translation in the Minkowski time does not commute with that in the Rindler time. 
It leads to an uncertainty inequality of the Minkowski and Rindler momenta and we see when this inequality is saturated.

The remainder of the paper is organized as follows: 
In section~\ref{sec:Mink-packet}, we review wave packets in Minkowski spacetime.
In section~\ref{sec:RindWave}, we briefly review QFT on Rindler space and introduce wave packets for Rindler observers.
In section~\ref{sec:Rindler-wave-in-Mink}, we investigate how Rindler wave packets appear to Minkowski observers. 
In section~\ref{sec:P(k)}, we compute the distribution of the Minkowski momentum for the Rindler wave packets. 
In sections~\ref{sec:<k>} and \ref{sec:Var-k}, we study the average and variance of the Minkowski momentum for the Rindler wave packets. 
In section~\ref{sec:uncert}, we see the Minkowski position–momentum uncertainty relation for the Rindler wave packets. 
Section~\ref{sec:concl} is devoted to conclusions.

%%%%%%%%%%%%%%%%%%%%%%%%%%%%%%%%%%%%%%%%%%%%%%%%%%%%%%%%%%%%%%
\section{Wave packets of scalar fields}
\label{sec:Mink-packet}
%%%%%%%%%%%%%%%%%%%%%%%%%%%%%%%%%%%%%%%%%%%%%%%%%%%%%%%%%%%%%%
We here summarize our notations and the basics of the wave packets. 
In this paper, we consider a free real massless scalar on a two-dimensional spacetime. 
We will use coordinates $(T, X)$ for the Minkowski spacetime and $(t,x)$ for the Rindler spacetime.  

The free massless scalar on the two-dimensional Minkowski spacetime can be expanded as
\begin{align}
\label{phi-Mink}
    \phi(T,X)=\int^\infty_{-\infty}\frac{dk}{\sqrt{4\pi\omega_k}}\left[a_k e^{-i(\omega_k T-k X)}+a_k^\dagger e^{+i(\omega_k T-k X)}\right],
\end{align}
where $\omega_k=|k|$. The commutation relations of $a_k, a_k^\dagger$ are
\begin{align}
[a_{k},a^{\dagger}_{k'}]=\delta(k-k'), \qquad 
[a_{k},a_{k'}]=[a^{\dagger}_{k},a^{\dagger}_{k'}]=0.
\end{align}
Generic one-particle states are given by
\begin{align}
    \ket{\psi}=\int_{-\infty}^{\infty}dk\,c(k) a^\dagger_k \ket{0}_M
\end{align}
with an arbitrary function $c(k)$,
where $\ket{0}_M$ is the Minkowski vacuum ($a_k \ket{0}_M=0)$.
The wave functions for these $\ket{\psi}$ are given by 
\begin{align}
\label{wf-mink}
    \psi(T,X):=\tensor[_M]{\bra{0}}{}\phi(T,X)\ket{\psi}=\int_{-\infty}^{\infty}\frac{dk}{\sqrt{4\pi\omega_k}}\,c(k)e^{-i(\omega_k T-k X)}.
\end{align}
For the momentum eigenstates $a^\dagger_k \ket{0}_M$, the wave functions are $\frac{1}{\sqrt{4\pi\omega_k}}e^{-i(\omega_k T-k X)}$ which are plane waves (up to the factor $1/\sqrt{4\pi\omega_k}$) and not localized on the position space.

The momentum eigenstates are inappropriate for discussing the position-dependent blue-shift of momenta. 
We have to consider wave-packet states sufficiently localized in the position space.
Let the wave functions of such wave-packet states have the form
\begin{align}
\label{wf-fk}
    \psi(T,X)=\int_{-\infty}^{\infty}\frac{dk}{\sqrt{4\pi\omega_k}}\,f_{k_0,\Delta k}(k) e^{-i \omega_k(T-T_0)+ik (X-X_0)}.
\end{align}
We suppose that $f_{k_0,\Delta k}(k)$ is localized at $k=k_0>0$ with width $\Delta k$ and simultaneously $\psi(T,X)$ is localized sufficiently around $X=X_0$ when $T=T_0$. Later, we will give an example of the case that $f_{k_0,\Delta k}(k)$ is a Gaussian function. 

The state $\ket{\psi}$ with wave function \eqref{wf-fk} is given by $\ket{\psi}=a^\dagger_\psi \ket{0}_M$ with
\begin{align}
\label{def:a-psi}
    a^\dagger_\psi=\int_{-\infty}^{\infty}dk\,f_{k_0,\Delta k}(k) a^\dagger_k e^{i \omega_k T_0-ik X_0}.
\end{align}
We fix the normalization of $f_{k_0,\Delta k}$ so that we have $\braket{\psi|\psi}=1$. 
The normalization condition is written as
\begin{align}
\label{fk-norm}
    \int_{-\infty}^{\infty}dk\,|f_{k_0,\Delta k}(k)|^2=1,
\end{align}
and it is equivalent to  
\begin{align}
    [a_\psi, a^\dagger_\psi]=1.
\end{align}
The normalization condition is also the same as the normalization of the relativistic density (or norm in the Klein-Gordon inner product)
\begin{align}
    \int dX\, j_0=1,
\end{align}
where $j_\mu$ is the relativistic density current
\begin{align}
    j_\mu:=i(\psi^\ast \partial_\mu \psi -(\partial_\mu \psi^\ast)\psi).
\end{align}

%%%%%%%%%%
\paragraph{The Gaussian wave packets:}
The Gaussian wave packets are obtained by taking the following $f_{k_0,\Delta k}$:
\begin{align}
    f_{k_0,\Delta k}(k)=\frac{1}{(2\pi \Delta k^2)^{\frac{1}{4}}}e^{-\frac{1}{4}\left(\frac{k-k_0}{\Delta k}\right)^2}.
\end{align}
The following integral gives the wave function:
\begin{align}
    \psi(T,X)&=\int_{-\infty}^{\infty}\frac{dk}{\sqrt{4\pi\omega_k}}\,f_{k_0,\Delta k}(k) e^{-i \omega_k(T-T_0)+ik (X-X_0)}
    \nn
    &=\frac{1}{(2\pi \Delta k^2)^{\frac{1}{4}}}\int_{-\infty}^{\infty}\frac{dk}{\sqrt{4\pi\omega_k}}\,e^{-\frac{1}{4}\left(\frac{k-k_0}{\Delta k}\right)^2-i \omega_k(T-T_0)+ik (X-X_0)}.
\end{align}
Using the saddle point approximation assuming $k_0/\Delta k > 2\Delta k|(T-T_0)-(X-X_0)|$, we obtain 
\begin{align}
\label{psi(T,X)_k}
   \psi(T,X)\simeq 
   \frac{e^{-\Delta k^2 (U-U_0)^2-i k_0 U_0}}{(2\pi)^{\frac{1}{4}}\sqrt{k_0/\Delta k-2i\Delta k (U-U_0)}},
\end{align}
where $U\equiv T-X$, $U_0\equiv T_0-X_0$.
Thus, (the absolute square of) the wave function is localized around $U=U_0$ (i.e., the peak of the wave moves along $X=T-T_0+X_0$) with the width $1/(2\Delta k)$.

%%%%%%%%%%%%%%%%%

%%%%%%%%%%%%%%%%%%%%%%%%%%%%%%%%%%%%%%%%%%%%%%%%%%%%%%%%%%%%%%
%%%%%%%%%%%%%%%%%%%%%%%%%%%%%%%%%%%%%%%%%%%%%%%%%%%%%%%%%%%%%%
\section{Wave packets for Rindler observers}\label{sec:RindWave}
%%%%%%%%%%%%%%%%%%%%%%%%%%%%%%%%%%%%%%%%%%%%%%%%%%%%%%%%%%%%%%
In this section, we consider the wave-packet states for the Rindler observers. 

%%%%%%%%%%%%%%%%%%%%%%%%%%%%%%%%%%%%%%%%%%%%%%%%%%%%%%%%%%%%%%
\subsection{Rindler spacetime and Rindler quantization}
In this subsection, we briefly introduce the two-dimensional  Rindler spacetime \cite{Rindler:1966zz} and the Rindler quantization of the massless scalar field.

For the right Rindler wedge $|T| \leq X$, we introduce the Rindler coordinates $(t, x)$ as 
\begin{align}
    T \pm X= \pm a^{-1}e^{\pm a(t \pm x)}.
    \label{Rind-coords}
\end{align}
Using the coordinates $U=T-X, V=T+X$ and $u=t-x, v=t+x$, the above coordinate transformation is given by
\begin{align}
    aU=-e^{-a u}, \quad  aV=e^{a v}.
\end{align}
The metric reads
\begin{align}
    ds^2=-dT^2+dX^2=e^{2 a x}(-dt^2+dx^2). 
\end{align}
The trajectory at constant $x$ corresponds to a uniformly accelerating observer with the proper acceleration $a e^{-a x}$ in the Minkowski spacetime.

This metric can also be obtained by the near-horizon limit of the Schwarzschild black hole geometry as reviewed in appendix~\ref{app:Rind}.
The black hole with mass $M$ leads to the Rindler space with $a=1/(4 G_N M)$ where $G_N$ is the Newton constant.

Let us consider the classical light signal emitted at a spacetime point $P=(T_0, X_0)$ with momentum $k^\mu=(|k|,k)$ in the Minkowski frame. 
In the Rindler frame, the point $P$ is at $(t_0,x_0)$ which is related to $(T_0, X_0)$ as \eqref{Rind-coords}. 
The frequency of the signal for the Minkowski observer is $\omega_M=|k|$. 
For the Rindler observer at $P$, the frequency per the Rindler time\footnote{This is not the proper time except at the point $x=0$. The frequency $\omega_p$ in \eqref{phi-Rind} is the one associated with the coordinate $t$ and not the proper time.} is given by 
\begin{align}
    \omega_R=\frac{dU}{du}\omega_M=e^{-au}\omega_M
\end{align}
for the outgoing signal ($k>0$).
This is the classical relation between the energy of a particle for Minkowski and Rindler observers. 
The red-shift factor $e^{-au}$ depends on the position $u$ of the particle. 
As mentioned in the previous section, the momentum eigenstates are inappropriate to discuss the red/blue-shift in quantum theories, and we should consider wave-packet states.

Before discussing wave packets in the Rindler spacetime, let us give a brief summary of the Rindler quantization of the free massless scalar as in section~\ref{sec:Mink-packet}.
The massless Klein-Gordon equation in the Rindler coordinates is 
\begin{align}
    (-\partial_t^2+\partial_x^2)\phi=0.
\end{align}
This is the same form as in the Minkowski space because our theory has a conformal symmetry. 
Thus, the positive frequency modes in the Rindler time $t$ take the plane wave form $e^{-i(\omega_p t- p x)}$ as in the Minkowski coordinates. 
Using the Rindler modes, we expand $\phi$ as
\begin{align}
\label{phi-Rind}
    \phi(t,x)=\int^\infty_{-\infty}\frac{dp}{\sqrt{4\pi\omega_p}}\left[b_p e^{-i(\omega_p t-p x)}+b_p^\dagger e^{+i(\omega_p t-p x)}\right].
\end{align}
The coefficients $b_p, b^\dagger_p$ satisfy 
\begin{align}
[b_{p},b^{\dagger}_{p'}]=\delta(p-p'), \qquad 
[b_{p},b_{p'}]=[b^{\dagger}_{p},b^{\dagger}_{p'}]=0.
\end{align}

The Rindler annihilation (creation) operators $b_p, b^\dagger_p$ are written in terms of the Minkowski operators $a_k, a^\dagger_k$ as
\begin{align}
    b_{p}=\int^{\infty}_{-\infty} dk \left[ \alpha_{p,k}^\ast a_k- \beta^\ast_{p,k} a_k^\dagger\right], \qquad b^\dagger_{p}=\int^{\infty}_{-\infty} dk \left[ \alpha_{p,k} a_k^\dagger- \beta_{p,k} a_k\right],
    \label{Bogo-b-a}
\end{align}
as summarized in appendix~\ref{app:Bogo}. 
The explicit expression of the Bogoliubov coefficients $\alpha_{p,k}, \beta_{p,k}$ is given by \eqref{Bogo-a}, \eqref{Bogo-b}, and for $p>0, k>0$, they become
\begin{align}
    \alpha_{p,k}=
    \frac{e^{\frac{\pi p}{2a}}}{2\pi a}\sqrt{\frac{p}{k}}
    \left(\frac{a}{k}\right)^{\frac{i p}{a}}\Gamma(i p/a),\quad
    \beta_{p,k}=
    \frac{-e^{-\frac{\pi p}{2a}}}{2\pi a}\sqrt{\frac{p}{k}}
    \left(\frac{a}{k}\right)^{\frac{i p}{a}}\Gamma(i p/a).
\end{align}

The Rindler vacuum $\ket{0}_R$ is defined as the state annihilated all $b_p$ as $b_p \ket{0}_R=0$. 
The Minkowski vacuum $\ket{0}_M$ is different from the Rindler one $\ket{0}_R$, and $\ket{0}_M$ behaves as the thermal state with temperature $a/2\pi$ for the Rindler particles.

Classically, a light with Rindler momentum $p>0$ along the line $u=u_0$ has the momentum $k=e^{a u_0} p$ in the Minkowski frame. 
We will study how this classical relation changes for the wave packets. 

%%%%%%%%%%%%%%%%%%%%%%%%%%%%%%%%%%%%%%%%%%%%%%%%%%%%%%%%%%%%%%
\subsection{Wave packet in Rindler spacetime}\label{subsec:wp-Rindler}
%%%%%%%%%%%%%%%%%%%%%%%%%%%%%%%%%%%%%%%%%%%%%%%%%%%%%%%%%%%%%%
We consider wave-packet states for the Rindler observer.
In the Minkowski spacetime, we introduce the creation operator $a^\dagger_\psi$ corresponding to the wave-packet wave function in \eqref{def:a-psi}.
We define a similar creation operator $b_\psi^\dagger$ as
\begin{align}
\label{def:b-psi}
    b^\dagger_\psi=\int_{-\infty}^{\infty}dp\,f_{p_0,\Delta p}(p) b^\dagger_p e^{i \omega_p t_0-ip x_0}.
\end{align}
We demand that $f_{p_0,\Delta p}$ satisfies the normalization condition
\begin{align}
\label{fp-norm}
    \int_{-\infty}^{\infty}dp\,|f_{p_0,\Delta p}(p)|^2=1
\end{align}
as similar to \eqref{fp-norm}.
If $b^\dagger_\psi$ acts on the Rindler vacuum $\ket{0}_R$, the analysis is the same as the Minkowski case. 
In this paper, we suppose that $f_{p_0,\Delta p}(p)$ has the support only for $p>0$, i.e., we consider outgoing wave packets because the analysis of the incoming case is parallel. 
Then, the normalization condition becomes
\begin{align}
\label{fp-norm2}
    \int_{0}^{\infty}dp\,|f_{p_0,\Delta p}(p)|^2=1.
\end{align}

Instead of $\ket{0}_R$, we regard $\ket{0}_M$ as the base state and consider the excitation around it. That is, we consider the states obtained by acting $b_\psi^\dagger$ on the Minkowski vacuum $\ket{0}_M$ as
\begin{align}
\label{def:psiR}
    \ket{\psi_R}:= \frac{1}{\sqrt{\cN}}b^\dagger_\psi \ket{0}_M,
\end{align}
where $\sqrt{\cN}$ is a normalization factor. Indeed, $b^\dagger_\psi \ket{0}_M$ is generally not normalized unlike $b^\dagger_\psi \ket{0}_R$.
We have
\begin{align}
    \braket{\psi_R|\psi_R}&=\cN^{-1} \tensor[_M]{\bra{0}}{} b_\psi b^\dagger_\psi \ket{0}_M
    \nn
    &=\cN^{-1}
    \int_{-\infty}^{\infty}dp dp'\,f^\ast_{p_0,\Delta p}(p)f_{p_0,\Delta p}(p') e^{-i (\omega_p-\omega_{p'}) t_0 +i(p-p') x_0} \tensor[_M]{\bra{0}}{}b_{p} b^\dagger_{p'}\ket{0}_M.
\end{align}
We focus on the case that $f_{p_0,\Delta p}(p)$ has the support only for $p>0$. 
Then, using \eqref{intk-aa}, we obtain
\begin{align}
    \braket{\psi_R|\psi_R}&=\cN^{-1}\int_{0}^{\infty}dp\,|f_{p_0,\Delta p}(p)|^2\frac{e^{\frac{2\pi \omega_p}{a}}}{e^{\frac{2\pi \omega_p}{a}}-1}.
    \label{psi-normalization}
\end{align}
We thus take the normalization factor as
\begin{align}
    \label{normal-fact}
    \cN=\int_{0}^{\infty}dp\,|f_{p_0,\Delta p}(p)|^2\frac{e^{\frac{2\pi \omega_p}{a}}}{e^{\frac{2\pi \omega_p}{a}}-1}.
\end{align}

The smearing function $f_{p_0,\Delta p}(p)$ is localized around $p=p_0$. Thus, if we consider a state $b^\dagger_\psi \ket{0}_R$, it is a wave-packet state for the Rindler particles, and the Rindler momentum is almost $p_0$. Explicitly, the expectation value of the Rindler momentum is
\begin{align}
   \int^\infty_{-\infty} dp \tensor[_R]{\bra{0}}{}b_{\psi} (p  b^\dagger_{p} b_p )b_{\psi}^\dagger \ket{0}_R=\int^\infty_0 dp\, p |f_{p_0,\Delta p}(p)|^2.
\end{align}
Hence, if $f_{p_0,\Delta p}(p)$ is well-localized around $p=p_0$ and normalized as \eqref{fp-norm2}, the expectation value is almost $p_0$.

However, we are interested in $\ket{\psi_R}= \frac{1}{\sqrt{\cN}}b^\dagger_\psi \ket{0}_M$, not $b_{\psi}^\dagger \ket{0}_R$. 
One might think that the Rindler momentum of $\ket{\psi_R}$ is not approximated by $p_0$. 
Nevertheless, if the peak $p_0$ of the smearing function is much larger than the Unruh temperature $a/(2\pi)$ (i.e. $p_0 \gg a$), we can ignore Unruh thermal fluctuations of $\ket{0}_M$, and the expectation value of the momentum\footnote{More precisely, the increment of the momentum from $\ket{0}_M$ due to the excitation by $b^\dagger_{\psi}$ is approximated by $p_0$.} is approximated by $p_0$. 
Since we aim to discuss the blue-shift of the momentum of the wave packets, the wave-packet states $\ket{\psi_R}$ are desired to have a well-localized momentum distribution. 
We thus do not consider the case that the peak $p_0$ is hidden by thermal fluctuations at the Unruh temperature, and we will focus on the case $p_0 \gg a$.\footnote{For the analysis of the Hawking radiation using the wave packet, it is natural to consider $p_0 \sim \cO(a)$, and then the situation is different.}
In this situation, the normalization factor \eqref{normal-fact} is almost 1 as
\begin{align}
    \cN
    \simeq \int_{0}^{\infty}dp\,|f_{p_0,\Delta p}(p)|^2\frac{e^{\frac{2\pi p_0}{a}}}{e^{\frac{2\pi p_0}{a}}-1} \simeq \int_{0}^{\infty}dp\,|f_{p_0,\Delta p}(p)|^2=1.
    \label{Nsim1}
\end{align}

We define the wave function for the state $\ket{\psi_R}$ in the same way as \eqref{wf-mink}, 
\begin{align}
\label{wf-rind} 
    \psi_R(t,x):=\tensor[_M]{\bra{0}}{}\phi(t,x)\ket{\psi_R},
\end{align}
where note that we have used $\ket{0}_M$, not $\ket{0}_R$ because we use $\ket{0}_M$ as the base state in the definition of $\ket{\psi_R}$ in \eqref{def:psiR}.
We can find that the wave function takes the form of
\begin{align}
    \psi_R(t,x)=\frac{1}{\sqrt{\cN}}\int^\infty_0
\frac{dp}{\sqrt{4\pi p}}
f_{p_0,\Delta p}(p)\frac{e^{\frac{2\pi p}{a}}}{e^{\frac{2\pi p}{a}}-1}e^{-ip(u-u_0)}.
\end{align}
Using the same approximation in \eqref{Nsim1} taking $p_0 \gg a$, we obtain
\begin{align}
    \psi_R(t,x)\simeq \int^\infty_0
\frac{dp}{\sqrt{4\pi p}}
f_{p_0,\Delta p}(p)e^{-ip(u-u_0)}.
\end{align}
This also shows that we can ignore the Unruh thermal effect because the above integral is the same one for the Minkowski wave function in \eqref{wf-fk} and thus the one for the Rindler vacuum $\ket{0}_R$ instead of $\ket{0}_M$.

Thus, for the Gaussian wave function
\begin{align}
\label{Gauss_Rind}
    f_{p_0,\Delta p}(p)=\frac{1}{(2\pi \Delta p^2)^{\frac{1}{4}}}e^{-\frac{1}{4}\left(\frac{p-p_0}{\Delta p}\right)^2},
\end{align}
we obtain
\begin{align}
   \psi_R(t,x)\simeq 
   \frac{e^{-\Delta p^2 (u-u_0)^2-i p_0 u_0}}{(2\pi)^{\frac{1}{4}}\sqrt{p_0/\Delta p-2i\Delta p (u-u_0)}}
\end{align}
as similar to \eqref{psi(T,X)_k}.
It shows that $\psi_R(u=t-x)$ has a peak around $u=u_0$ and the width $\Delta u$ is related to $\Delta p$ as
\begin{align}
\label{du-dp}
   \Delta u \simeq 1/(2\Delta p). 
\end{align}

%%%%%%%%%%%%%%%%%%%%
\section{Rindler wave packets in Minkowski}
\label{sec:Rindler-wave-in-Mink}
%%%%%%%%%%%%%%%%%%%%
We study what the wave-packet states $ \ket{\psi_R}= \frac{1}{\sqrt{\cN}}b^\dagger_\psi \ket{0}_M$ defined in \eqref{def:psiR} in the previous section appear to the Minkowski observer. 
%%%%%%%%%%%%%%%%%%%%
\subsection{Minkowski momentum of the Rindler wave packets}
\label{sec:P(k)}
%%%%%%%%%%%%%%%%%%%%
We note that the states $\ket{\psi_R}$ are one-particle states for the Minkowski modes as
\begin{align}
    \ket{\psi_R}= \frac{1}{\sqrt{\cN}}\int_{-\infty}^{\infty}dk\,\left(\int_{-\infty}^{\infty}dp\,f_{p_0,\Delta p}(p) \alpha_{p,k} e^{i \omega_p t_0 -ip x_0}\right)a_k^\dagger \ket{0}_M,
\end{align}
where we have used the Bogoliubov transformation \eqref{Bogo-b-a}.

The wave function \eqref{wf-rind} is expressed in terms of the Minkowski coordinates as
\begin{align}
    \psi_R(T,X)=\tensor[_M]{\bra{0}}{}\phi(T,X)\ket{\psi_R}=\int_{-\infty}^{\infty}\frac{dk}{\sqrt{4\pi\omega_k}}\,c_R(k)e^{-i(\omega_k T-k X)},
\end{align}
where $c_R(k)$ is defined as
\begin{align}
\label{def:cr}
    c_R(k):=\frac{1}{\sqrt{\cN}}\int_{-\infty}^{\infty}dp\,f_{p_0,\Delta p}(p) \alpha_{p,k} e^{i \omega_p t_0 -ip x_0}.
\end{align}
$\psi_R$ has a peak at $u=u_0$ and width $\Delta u$ in the Rindler coordinates $u$.\footnote{Since $\alpha_{p,k}$ is singular at $p=0$, we should avoid this singularity. Here we suppose that the integration contour of $p$ is slightly shifted into the lower half‑plane, just below the real axis.
\label{foot1}} 
In the Minkowski coordinates $U=-a^{-1}e^{-a u}$, $\psi_R$ has a peak at $U=U_0:=-a^{-1}e^{-a u_0}$ with the width 
\begin{align}
    \Delta U=[-a^{-1}e^{-a u}]^{u=u_0+\Delta u/2}_{u=u_0-\Delta u/2}=2a^{-1}e^{-a u_0} \sinh\left(\frac{a \Delta u}{2}\right).
\end{align}
For the Gaussian function \eqref{Gauss_Rind}, we have $   \Delta u \sim 1/(2\Delta p)$. 
If $\Delta p \gg a$, we have $\Delta u\ll a^{-1}$ and 
\begin{align}
    \Delta U \sim e^{-a u_0} \Delta u \sim e^{-a u_0} (\Delta p)^{-1}.
    \label{DeltaU-Deltap-1}
\end{align}
On the other hand, if $\Delta p \lesssim a$,
\begin{align}
    \Delta U \sim 2a^{-1}e^{-a u_0} \sinh\left(\frac{a}{4\Delta p}\right).
    \label{DeltaU-Deltap-2}
\end{align}
Note that $\sinh\left(\frac{a}{4\Delta p}\right)$ is not small in this case but $\Delta U$ might be small near the horizon where $u_0$ is positively large, i.e., $U_0$ is small.

We now consider the distribution of the Minkowski momentum $k$ for the state $\ket{\psi_R}$. 
Let $P(k)$ be the probability density of the Minkowski momentum $k$ for $\ket{\psi_R}$ as
\begin{align}
\label{def:P(k)}
    P(k):=\bra{\psi_R}a^\dagger_k a_k \ket{\psi_R}.
\end{align}
It is given by
\begin{align}
    P(k)=|[a_k,b^\dagger_\psi]|^2=\left|\frac{1}{\sqrt{\cN}}\int_{-\infty}^{\infty}dp\,f_{p_0,\Delta p}(p) \alpha_{p,k} e^{i \omega_p t_0 -ip x_0}\right|^2=|c_R(k)|^2.
\end{align}
Note that we have
\begin{align}
    \int^\infty_{-\infty} dk P(k)=\braket{\psi_R|\psi_R}=1.
\end{align}
because $\ket{\psi}$ is a one-particle state for the Minkowski particle.

The mean value of $k^n$ under the probability density $P(k)$ is given by
\begin{align}
    \braket{k^n}=\int^\infty_{-\infty} dk\, k^n P(k).
\end{align}
As mentioned above, we focus on the case that $f_{p_0,\Delta p}(p)$ has the support only for $p>0$. Then, $P(k)$ vanishes for $k\leq0$ because of the explicit form of the Bogoliubov coefficient $\alpha_{p,k}$ in \eqref{Bogo-a} (due to the chirality of a two-dimensional massless scalar). 
It leads to
\begin{align}
    \braket{k^n}&=\int^\infty_{0} dk\, k^n P(k)=a^{n+1}e^{(n+1)a u_0}\int^\infty_{0} d\tk\, \tk^n P(a e^{au_0}\tk)
    \nn
    &=\frac{a^{n+1}e^{(n+1)a u_0}}{\cN}\int^\infty_{0} d\tk \, \tk^n \left|\int_{0}^{\infty}dp\,f_{p_0,\Delta p}(p) \alpha_{p,a e^{au_0}\tk} e^{i p u_0}\right|^2,
    \label{kn}
\end{align}
where $u_0=t_0-x_0$, and we have rescaled the integration variables as $k=a e^{au_0}\tk$.
From \eqref{Bogo-a}, we also have
\begin{align}
    \alpha_{p,a e^{au_0}\tk}\, e^{i p u_0}=\frac{e^{-\frac{au_0}{2}+\frac{\pi p}{2a}}}{2\pi a} \tk^{-\frac{i p}{a}-\frac{1}{2}}
    \sqrt{\frac{p}{a}}
   \Gamma\left(\frac{i p}{a}\right).
\end{align}
We thus have
\begin{align}
\label{barkn}
    \braket{k^n}=\frac{a^{n}e^{n a u_0}}{\cN}\int^\infty_{0} d\tk \tk^{n-1}\, |I_+(\tk)|^2
\end{align}
with 
\begin{align}
\label{def:I+}
I_+(\tk):=\sqrt{a}\int_{0}^{\infty}dp\,f_{p_0,\Delta p}(p)  \frac{e^{\frac{\pi p}{2a}}}{2\pi a}\tk^{-\frac{i p}{a}}
\sqrt{\frac{p}{a}}\,
   \Gamma\!\left(\frac{i p}{a}\right),
\end{align}
where we note that $I_+(\tk)$ is dimensionless.\footnote{$c_R(k)$ in \eqref{def:cr} is related to $I_+(\tk)$ as $c_R(k=a e^{a u_0}\tk)=\frac{e^{-a u_0/2}}{\sqrt{\cN a \tk}}I_+(\tk)$.
In addition, as noted in footnote~\ref{foot1}, we supposed that the integration contour of $p$ is slightly shifted into the lower half‑plane, just below the real axis.
\label{foot2}} 
Therefore, if we obtain this $I_+(\tk)$, we can compute $\braket{k^n}$ from \eqref{barkn}.
We rewrite \eqref{barkn}, factoring out $u_0$-dependence, as
\begin{align}
    \braket{k^n}=a^{n}e^{n a u_0}\braket{\tk^n},
    \qquad
    \braket{\tk^n}:=
    \int^\infty_{0} d\tk \tk^{n}\, \tilde{P}(\tk),
\end{align}
where $\tilde{P}(\tk)$ is the probability distribution for the rescaled momentum $\tk$, and is given by
\begin{align}
    \tilde{P}(\tk):=a e^{a u_0}P(a e^{a u_0}\tk)=\frac{1}{\cN}\, \frac{|I_+(\tk)|^2}{\tk}.
    \label{def:tilP}
\end{align}
This $\tilde{P}(\tk)$ is independent of $u_0$.

\begin{figure}[H]
\begin{center}
\includegraphics[width=0.8\columnwidth]{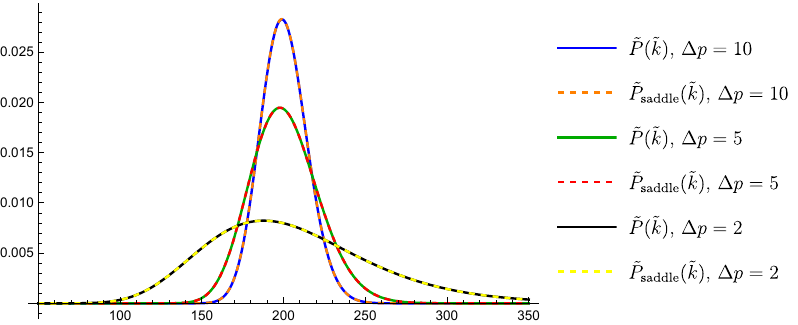}
  \vspace{-1em}
  \end{center}
  %\vspace{-1em}
  \caption{Plots of $\tilde{P}(\tk)$ and $\tilde{P}_\text{saddle}(\tk)$ for various $\Delta p$ with $\Delta p > a$.  We set $\Delta p=10,5,2$ with $p_0=200, a=1$. These plots show that $\tilde{P}_\text{saddle}(\tk)$ excellently agrees with $\tilde{P}(\tk)$.}
  \label{fig:p0-200_a-1}
\end{figure}

\begin{figure}[H]
\begin{center}
\includegraphics[width=0.8\columnwidth]{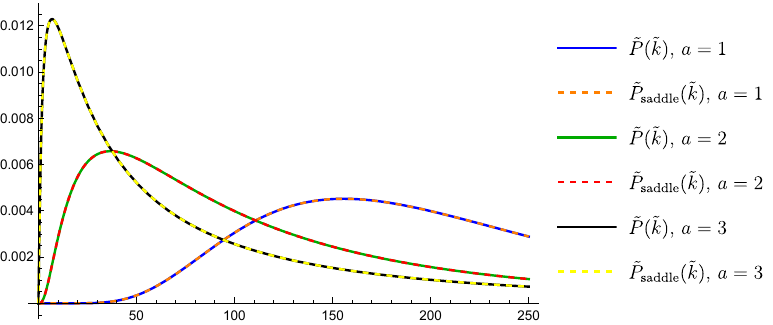}
  \vspace{-1em}
  \end{center}
  %\vspace{-1em}
  \caption{Plots of $\tilde{P}(\tk)$ and $\tilde{P}_\text{saddle}(\tk)$ for $\Delta p\leq a$. In the plots, we set $a=1,2,3$ with $p_0=200, \Delta p=1$. These plots also show that $\tilde{P}_\text{saddle}(\tk)$ excellently agrees with $\tilde{P}(\tk)$.}
  \label{fig:p0-200_dp-1}
\end{figure}

We will evaluate $\tilde{P}(\tk)$ using the saddle point approximation for the Gaussian wave packets in appendix~\ref{app:LambertW}.
Here we just show the result.
For the Gaussian wave packets
\begin{align}
    f_{p_0,\Delta p}(p)=\frac{1}{(2\pi \Delta p^2)^{\frac{1}{4}}}e^{-\frac{1}{4}\left(\frac{p-p_0}{\Delta p}\right)^2},
    \label{gaussian}
\end{align}
supposing $p_0-\Delta p \gg a$, the saddle point approximation leads to
\begin{align}
|I_+(\tk)|&\simeq |I^\text{saddle}_+(\tk)|:=\frac{e^{\mathrm{Re}\left[\frac{p_\ast^2-p_0^2}{4\Delta p^2}-\frac{ip_\ast}{a}\right]}}{(2\pi)^{1/4}\left|\frac{a}{2\Delta p}-\frac{i \Delta p}{p_\ast}\right|^{\frac{1}{2}}},
\end{align}
where complex $p_\ast$ depends on $\tk$ as
\begin{align}
\label{main:past}
    p_\ast=-\frac{2i \Delta p^2}{a}W_{n}\left(\frac{i a^2 \tk}{2\Delta p^2}e^{\frac{i a p_0}{2\Delta p^2}}\right).
\end{align}
Here, $W_{n}(z)$ is the Lambert $W$ function (product logarithm), i.e., the inverse function of $z=w e^w$, which is expressed as $w=W_n(z)$. 
The label $n$ denotes the branch and is given by
\begin{align}
    n=\left\lceil \frac{\frac{a p_0}{2\Delta p^2}-\frac{\pi}{2}}{2\pi} \right\rceil,
    \label{branch-n}
\end{align}
where $\lceil \cdot \rceil$ is the ceiling function: $\lceil x\rceil:=\min\left\{n \in \mathbb{Z}\,|\, x \leq n\right\}$.
Using this saddle point approximation, $\tilde{P}(\tk)$ defined in \eqref{def:tilP} is given by
\begin{align}
    \tilde{P}(\tk)\simeq \tilde{P}_\text{saddle}(\tk):=\frac{e^{\mathrm{Re}\left[\frac{p_\ast^2-p_0^2}{2\Delta p^2}-\frac{2ip_\ast}{a}\right]}}{(2\pi)^{1/2}\,\tk\left|\frac{a}{2\Delta p}-\frac{i \Delta p}{p_\ast}\right|},
    \label{eq:saddleP}
\end{align}
with $p_\ast$ in \eqref{main:past}, where we have also used $\cN \simeq 1$ as shown in \eqref{Nsim1}.
The saddle point approximation shows good agreement as in Figs.~\ref{fig:p0-200_a-1}, \ref{fig:p0-200_dp-1}, regardless of whether $a$ is larger or smaller than $\Delta p$.

$\tilde{P}_\text{saddle}(\tk)$ has a well-localized function (like the Gaussian distribution) when $p_0 \gg \Delta p > a$, and the peak is around $p_0/a$ as in Fig.~\ref{fig:p0-200_a-1}, although the width of $\tilde{P}_\text{saddle}(\tk)$ becomes broad for small $\Delta p$.
However, when $p_0 \gg a  \geq \Delta p$, the decay of 
$\tilde{P}_\text{saddle}(\tk)$ in the direction of large 
 $\tk$ becomes slow, and the peak shifts toward $\tk=0$ from $p_0/a$ as in Fig.~\ref{fig:p0-200_dp-1}.

We can compute any expectation value of functions of the Minkowski momentum $k$ by using the obtained distribution
\begin{align}
\label{eq:P-tP}
     P(k)=a^{-1}e^{-a u_0}\tilde{P}_\text{saddle}(a^{-1}e^{-a u_0} k).
\end{align}

%%%%%%%%%%%%%%%%%%%%%%%
\subsection{Expectation values of Minkowski momentum}\label{sec:<k>}

We are now able to evaluate the expectation value $\langle k\rangle$ by using the obtained distribution \eqref{eq:P-tP}.
Although it is easy to do numerically, the analytical integration is not easy. 
We thus take a slightly different way. 

$\langle k^n\rangle$ in \eqref{barkn} for positive numbers $n$ is written as
\begin{align}
        \langle k^n\rangle=&\frac{a^n e^{nau_0}}{\mathcal{N}}\int_{-\infty}^{\infty}\frac{dp\,dp'}{4\pi^2a^2}\,f_{p_0,\Delta p}(p)f_{p_0,\Delta p}^*(p')\,e^{\frac{\pi}{2a}(p+p')}\,\sqrt{pp'}\,\Gamma\left(\frac{ip}{a}\right)\Gamma\left(\frac{-ip'}{a}\right)
        \nn
        &\qquad \qquad
       \times  \int_{0}^{\infty}d\tk \tk^{n-1}\tk^{-\frac{i}{a}(p-p')}.
\end{align}
We note that, although we have written the integration region of $p, p'$ to $(-\infty, \infty)$, the $p$‑contour is deformed slightly into the lower half‑plane and the $p'$‑contour into the upper half‑plane as noted in footnote~\ref{foot1}.
By setting $\tk=e^{a y}$, we have
\begin{align}
    \int_{0}^{\infty}d\tk \tk^{n-1}\tk^{-\frac{i}{a}(p-p')}=a \int_{-\infty}^{\infty}\!\! d  y\, e^{-i (p-p'+ in a)y}.
\end{align}
We now change the variables as $p \to p - i na /2$, $p' \to p' + i na /2$, and shift the contours of $p ,p'$ to the real axis where we assume that $f_{p_0,\Delta}(p)$ is holomorphic in the region $0\leq \mathrm{Im}\,p <n a/2$ so that we can safely change contours. Then the $y$-integration generates the delta function $\delta(p-p')$, and we obtain
\begin{align}
            \langle k^n\rangle=&\frac{a^{n} e^{nau_0}}{\mathcal{N}}\int_{-\infty}^{\infty}\!\frac{dp}{2\pi a}\,|f_{p_0,\Delta p}(p-i na/2)|^2 \,e^{\frac{\pi p}{a}}
           |p-i na/2|\,\left|\Gamma\left(\frac{ip}{a}+\frac{n}{2}\right)\right|^2.
\end{align}

In particular for $n=1, 2$, we obtain
\begin{align}
\label{<k>:f}
        \braket{k}&=\frac{e^{au_0}}{\mathcal{N}}\int_{-\infty}^{\infty}\!\!
        dp\,|f_{p_0,\Delta p}(p-i a/2)|^2 \,\frac{\sqrt{p^2+\frac{a^2}{4}}}{1+e^{-\frac{2\pi p}{a}}},
        \\
\label{<k2>:f}
         \braket{k^2}&=\frac{e^{2au_0}}{\mathcal{N}}\int_{-\infty}^{\infty}\!\!
        dp\,|f_{p_0,\Delta p}(p-i a)|^2 \,\frac{p \sqrt{p^2+a^2}}{1-e^{-\frac{2\pi p}{a}}}.
\end{align}
We note that these results reproduce the classical blue-shift 
\begin{align}
     \braket{k} \sim e^{a u_0} p_0, \quad \braket{k^2} \sim (e^{a u_0} p_0)^2,
\end{align}
if $a$ and $\Delta p$ are negligibly small so that the $p$-integration is dominated around $p=p_0$.

\begin{figure}[H]
\begin{center}
\includegraphics[width=0.5\columnwidth]{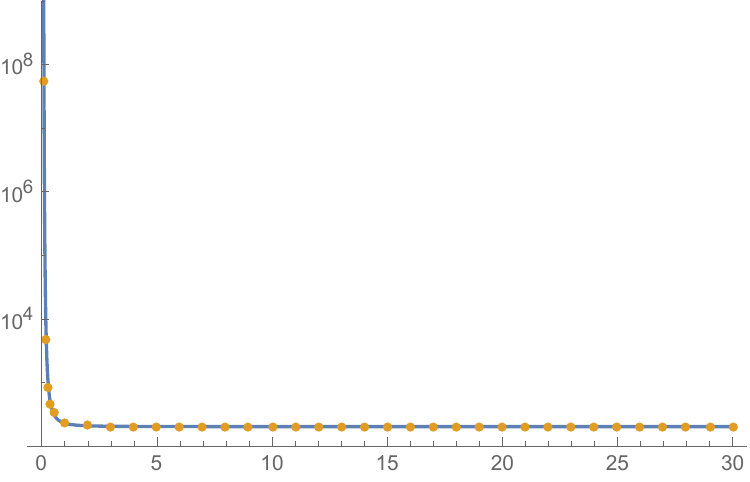}
  \vspace{-1em}
  \end{center}
  %\vspace{-1em}
  \caption{Plots of $\braket{k}$ (vertical) versus $\Delta p$ (horizontal) with the vertical axis on a logarithmic scale. The parameters are set as $p_0=200, a=1, u_0=0$.
  The blue curve is computed from \eqref{gaussk}. The orange dots are obtained from \eqref{eq:saddleP}. They are in good agreement. It justifies the use of \eqref{gaussk}.}
  \label{fig:<k>-dp}
\end{figure}

For the Gaussian wave packet \eqref{gaussian}, the above results \eqref{<k>:f} and \eqref{<k2>:f} lead to
\begin{align}
        \braket{k} &=\frac{e^{a u_0+\frac{a^2}{8\Delta p^2}}}{\sqrt{2 \pi} \Delta p} \int_{-\infty}^{\infty} dp\,e^{-\frac{1}{2}\left(\frac{p-p_0}{\Delta p}\right)^2}\frac{\sqrt{p^2+\frac{a^2}{4}}}{1+e^{-\frac{2 \pi p}{a}}} \,,
    \label{gaussk}
    \\
    \braket{k^2}&=\frac{e^{2au_0+\frac{a^2}{2\Delta p^2}}}{\sqrt{2 \pi} \Delta p} \int_{-\infty}^{\infty} dp\,e^{-\frac{1}{2}\left(\frac{p-p_0}{\Delta p}\right)^2}\frac{p\sqrt{p^2+a^2}}{1-e^{-\frac{2 \pi p}{a}}} \,,
    \label{gaussk2}
\end{align}
where we have used $\mathcal{N} \simeq 1$ as shown in \eqref{Nsim1}.
For a check of this result,  we illustrate that \eqref{gaussk} and \eqref{eq:saddleP} agree well in Fig.~\ref{fig:<k>-dp}. 

When $p_0 \gg a$ and $p_0 \gg \Delta p$, the integrals \eqref{gaussk} and \eqref{gaussk2} are approximated\footnote{The precise conditions for these approximation are $|f_1''(p_0)\Delta p^2| \ll f_1(p_0)$ with $f_1(p)=\sqrt{p^2+a^2/4}$ for $\braket{k}$ and $|f_2''(p_0)\Delta p^2| \ll f_2(p_0)$ with $f_2(p)=p\sqrt{p^2+a^2}$ for $\braket{k^2}$.}, by ignoring $a$ in the integrands, as
\begin{align}
        \braket{k} & \simeq\frac{e^{a u_0+\frac{a^2}{8\Delta p^2}}}{\sqrt{2 \pi} \Delta p} \int_{-\infty}^{\infty} dp\,e^{-\frac{1}{2}\left(\frac{p-p_0}{\Delta p}\right)^2} p=e^{a u_0+\frac{a^2}{8\Delta p^2}}p_0\,,
        \label{gaussk-app}
    \\
    \braket{k^2}& \simeq \frac{e^{2au_0+\frac{a^2}{2\Delta p^2}}}{\sqrt{2 \pi} \Delta p} \int_{-\infty}^{\infty} dp\,e^{-\frac{1}{2}\left(\frac{p-p_0}{\Delta p}\right)^2} p^2=e^{2au_0+\frac{a^2}{2\Delta p^2}}\left(1+\frac{\Delta p^2}{p_0^2}\right)p_0^2\,.
    \label{gaussk2-app}
\end{align}

The exponential factor $e^{au_0}$ in \eqref{gaussk-app} is the standard blue-shift factor. In addition, we have another enhancement factor $e^{\frac{a^2}{8\Delta p^2}}$ although it is small if  $a \ll \Delta p$. 
This enhancement factor is extremely large for a small $\Delta p$ such as $\Delta p \lesssim a$, and thus the trans-Planckian problem becomes more serious for such wave packets.
The spatial width $\Delta u (\sim 1/\Delta p)$ of such wave packets is larger than the length scale $1/a$ associated with the Unruh temperature as $\Delta u \gtrsim 1/a$. 
It indicates that the enhancement factor arises from the wave nature of the wave packet. 
Fig.~\ref{fig:p0-200_dp-1} indicates that $\tilde{P}(\tk)$ has a long tail, although the peak shifts toward the origin. The long tail effect overcomes the shifting effect of the peak, yielding a large value of $\braket{k}$.
Related to this property, in this parameter regime, the wave packet in the Minkowski coordinates departs radically from a Gaussian wave packet, as we shall see later.

\subsection{Variance of Minkowski momentum}\label{sec:Var-k}
We now consider the variance of Minkowski momentum 
\begin{equation}
    \Delta k=\sqrt{\braket{k^2}-\braket{k}^2} .
    \label{vark}
\end{equation}
To discuss the trans-Planckian problem, it is important to consider the variance $\Delta k$ because the validity of our EFT description requires that $\braket{k}+\Delta k$ be less than a cutoff scale.

We cannot minimize both $\Delta p$ and $\Delta k$ simultaneously. 
Indeed, we generally have the inequality \cite{Ho:2021sbi}
\begin{align}
     \Delta k\, \Delta p \geq  \frac{a}{2}\braket{k}
     \label{kp-uncern}
\end{align}
because the two operators 
$\hat{k}=i \partial_U$, $\hat{p}=i \partial_u=-a U \hat{k}$ does not commute with each other as
\begin{align}
    [\hat{k}, \hat{p}]=-i a  \hat{k}.
    \label{kp-commu}
\end{align}
In general, for two operators $\hat{A}, \hat{B}$, the Robertson inequality gives the uncertainty relation
\begin{align}
    \Delta A\, \Delta B \geq \frac{1}{2}\left|\langle [\hat{A}, \hat{B}]  \rangle\right|,
\end{align}
where $\Delta O:=\sqrt{\langle \hat{O}^2 \rangle - \langle \hat{O}\rangle^2}$.
Hence, the commutation relation \eqref{kp-commu} leads to the inequality \eqref{kp-uncern}.

\begin{figure}[htbp]
        \begin{minipage}{.5\textwidth}
            \centering
            \includegraphics[width=1\linewidth]{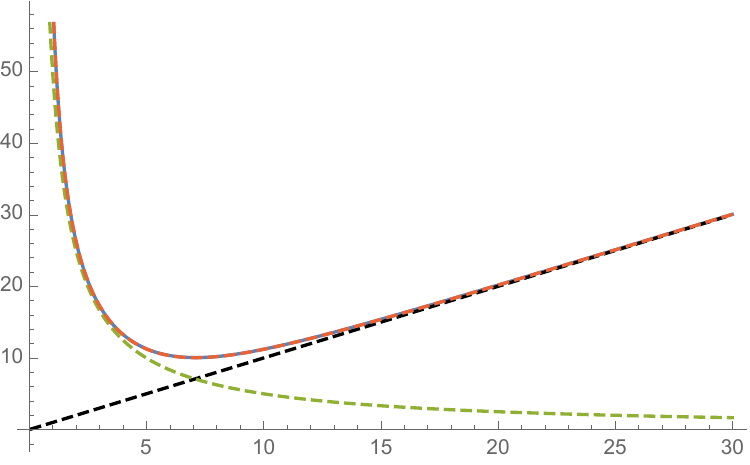}
        \end{minipage}
        \begin{minipage}{.5\textwidth}
            \centering
            \includegraphics[width=1\linewidth]{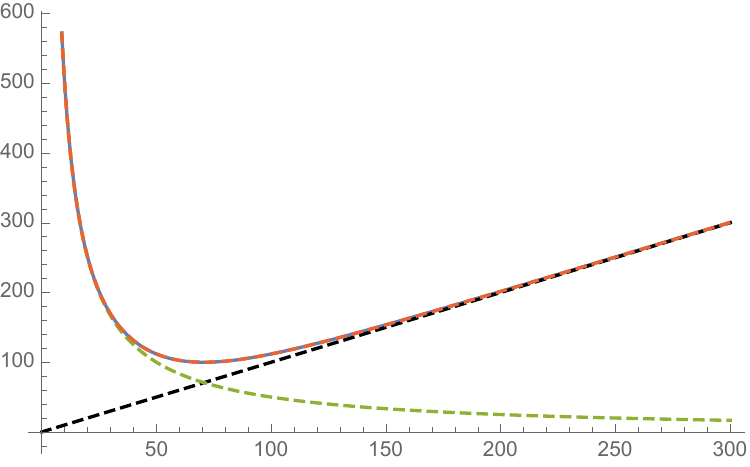}
        \end{minipage}
        \caption{Dependence of $\Delta k$ (the vertical axis) on $\Delta p$ (the horizontal one) for various $p_0$ values. In each panel, the blue curve is $\Delta k$ in \eqref{vark} computed from \eqref{gaussk} and \eqref{gaussk2} as a function of $\Delta p$, with $p_0=10^2$ (left) and $p_0=10^4$ (right), while $u_0=0, a=1$. The red, green, and black dashed curves are given by \eqref{eq:Deltak-app1}, \eqref{deltak1} and \eqref{deltak2} respectively. The red curves agree well with the blue ones. 
        These plots show that \eqref{eq:Deltak-app1} is a good approximation for $a \ll \Delta p \ll p_0$, and \eqref{deltak1} is good for $a \ll\Delta p < \sqrt{a p_0/2}$, and \eqref{deltak2} is good for $\sqrt{a p_0/2} < \Delta p \ll p_0$. }
        \label{fig:dp_vs_dk}
\end{figure}

We can compute $ \Delta k$ using \eqref{gaussk} and \eqref{gaussk2}.
The numerical plots for $\Delta k$ versus $ \Delta p$ are shown as blue curves in Fig.~\ref{fig:dp_vs_dk}.
The plots show that $\Delta k$ is proportional to $1/\Delta p$ (green curves) for some small $\Delta p$ region and proportional to $\Delta p$ (black curves) for some large $\Delta p$. 

We can understand these behaviors as follows.
The approximations \eqref{gaussk-app} and \eqref{gaussk2-app} lead to 
\begin{align}
    \Delta k \simeq e^{\frac{a^2}{4\Delta p^2}}\sqrt{1+\frac{\Delta p^2}{p_0^2}-e^{-\frac{a^2}{4\Delta p^2}}}e^{a u_0}p_0 \qquad (\text{for $a,  \Delta p \ll p_0$}).
    \label{eq:Deltak-app1}
\end{align}
In particular, for $a \ll \Delta p \ll p_0$, we have
\begin{align}
      \Delta k \simeq \sqrt{\frac{\Delta p^2}{p_0^2}+\frac{a^2}{4\Delta p^2}}e^{a u_0}p_0 \qquad (\text{for $a \ll \Delta p \ll p_0$}).
      \label{eq:Deltak-app2}
\end{align}
By further expanding it according to which term in the square root in \eqref{eq:Deltak-app2} dominates, i.e., $a/(2\Delta p) < \Delta p /p_0$ or $a/(2\Delta p) > \Delta p /p_0$, we obtain the following results
\begin{align}
     \Delta k \simeq e^{au_0}\,\frac{a \, p_0}{2\Delta p},\qquad (\text{for $a \ll \Delta p < \sqrt{a p_0/2} \ll p_0$}),
    \label{deltak1}
\end{align}
and 
\begin{align}
     \Delta k \simeq e^{au_0}\,\Delta p\qquad (\text{for $a \ll \sqrt{a p_0/2} <\Delta p \ll p_0$}).
    \label{deltak2}
\end{align}
Eq.~\eqref{deltak1} represents the green curves in Fig.~\ref{fig:dp_vs_dk}, and \eqref{deltak2} does the black lines.

Eq.~\eqref{deltak2}, that is, $\Delta k \simeq e^{au_0}\,\Delta p$, is an expected result from the Gaussian wave packet with the classical blue-shift factor because, as we have seen in Fig.~\ref{fig:p0-200_a-1}, the momentum distribution in $k$-space also takes a Gaussian form for $a \ll \sqrt{a p_0/2} <\Delta p \ll p_0$. Note also that the integer \eqref{branch-n} labeling the branch in the saddle point analysis becomes zero in this parameter regime, while it is not clear whether this integer \eqref{branch-n} has any physical meaning.

\begin{figure}[htbp]
        \begin{minipage}{.5\textwidth}
            \centering
            \includegraphics[width=1\linewidth]{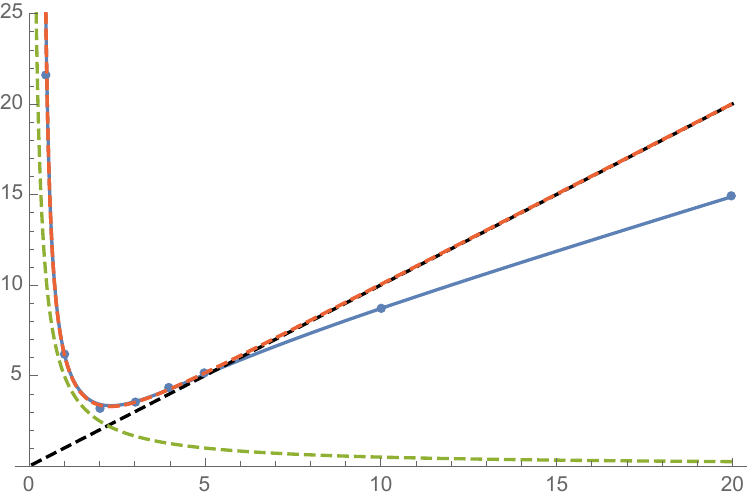}
        \end{minipage}
        \begin{minipage}{.5\textwidth}
            \centering
            \includegraphics[width=1\linewidth]{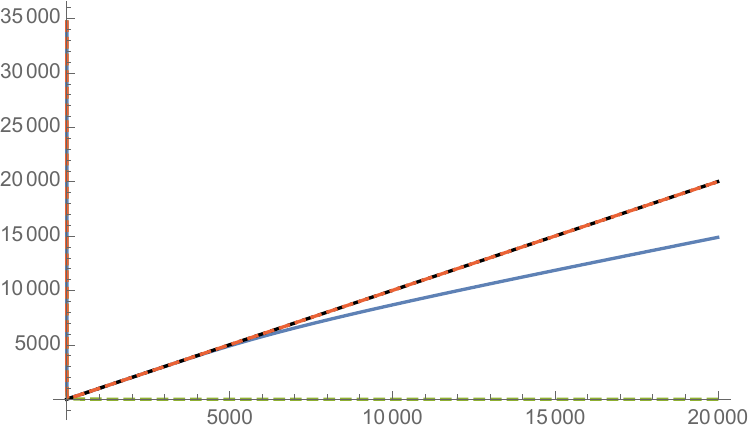}
        \end{minipage}
        \caption{Dependence of $\Delta k$ (vertical) on $\Delta p$ (horizontal) for $p_0=10$ (left) and $p_0=10^4$ (right) with $u_0=0, a=1$. The blue, red, green and black curves are obtained in the same way as in Fig.~\ref{fig:dp_vs_dk}. In the left plot ($p_0 =10$), the blue dots are obtained using the saddle point result \eqref{eq:saddleP}. It shows that \eqref{eq:Deltak-app1} is valid even for $p_0 /a=10$ if $\Delta p \ll p_0$. However, the approximation \eqref{deltak1} is not good because $a \ll\Delta p < \sqrt{a p_0/2}$ is not satisfied. 
        These plots also show that $\Delta k \simeq e^{au_0}\,\Delta p$ does not hold for large $\Delta p$ such as $\Delta p \lesssim p_0$.}
        \label{fig:dp_vs_dk_deviat}
\end{figure}

However, if $\Delta p \sim p_0$, we should include higher order corrections to the approximations \eqref{gaussk-app}, \eqref{gaussk2-app}.
Indeed, for large $\Delta p$ such that $\Delta p \lesssim p_0$, the approximation \eqref{deltak2} deviates from the numerical result of $\Delta k$ as in Fig.~\ref{fig:dp_vs_dk_deviat}.

For $a \ll \Delta p < \sqrt{a p_0/2} \ll p_0$, the $\Delta p$-dependence of $\Delta k$ drastically changes; $\Delta k$ is inversely proportional to $\Delta p$ as in \eqref{deltak1}, 
although the spatial width $\Delta u$ is still much smaller than $1/a$. 
In addition, the parameter regime $a \ll \Delta p < \sqrt{a p_0/2} \ll p_0$ is exceptional from the perspective of the uncertainty relation as follows.
In this parameter regime, we also have $\braket{k}  \simeq e^{a u_0}p_0$, and thus we obtain 
\begin{align}
    \Delta k\, \Delta p \simeq \frac{a}{2}\braket{k}.
    \label{kp-uncern-satu}
\end{align}

The result \eqref{kp-uncern-satu} implies that the uncertainty inequality \eqref{kp-uncern} is almost saturated for our Gaussian wave-packet states independent of the position of the center of the wave packet (or $u_0$), when $a \ll \Delta p < \sqrt{a p_0/2} \ll p_0$.
In other parameter regimes, we generally have $\Delta k\, \Delta p \gg  \frac{a}{2}\braket{k}$. Indeed, using \eqref{gaussk-app} and \eqref{eq:Deltak-app1}, we have
\begin{align*}
    \frac{\Delta k \Delta p}{\frac{a}{2}\braket{k}}\simeq 
    2e^{\frac{a^2}{8\Delta p^2}}\sqrt{1+\frac{\Delta p^2}{p_0^2}-e^{-\frac{a^2}{4\Delta p^2}}}\frac{\Delta p }a.
\end{align*}
It is generally much larger than 1. 
In this sense, the parameter regime $a \ll \Delta p < \sqrt{a p_0/2} \ll p_0$ is exceptional.

\begin{figure}[htbp]
\begin{center}
\includegraphics[width=0.5\columnwidth]{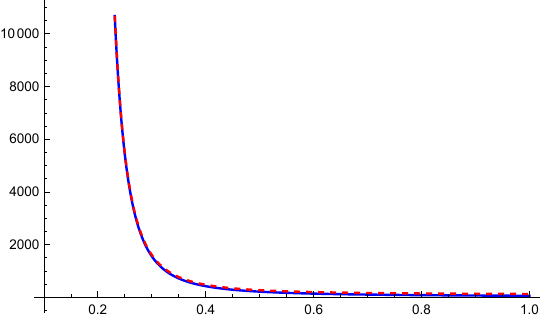}
  \vspace{-1em}
  \end{center}
  %\vspace{-1em}
  \caption{Plots of $\Delta k$ (vertical) versus $\Delta p$ (horizontal) for the case $\Delta p \ll a \ll p_0$. The parameters are set as $p_0=100, a=1, u_0=0$.
  The blue curve is computed from \eqref{gaussk} and \eqref{gaussk2}. The red dashed curve is obtained from \eqref{eq:dk-smalldp}. They are in good agreement.}
  \label{fig:dk-smalldp}
\end{figure}

If $\Delta p$ is much less than $a$ as $\Delta p \ll a \ll p_0$, \eqref{eq:Deltak-app1} leads to 
\begin{align} 
\label{eq:dk-smalldp}
    \Delta k \simeq e^{\frac{a^2}{4\Delta p^2}}e^{a u_0}p_0 \qquad (\text{for $\Delta p \ll a \ll p_0$}).
\end{align}
$\Delta k$ becomes extremely large due to the enhancement factor $e^{\frac{a^2}{4\Delta p^2}}$ in addition to the blue-shift factor $e^{a u_0}$ (see Fig.~\ref{fig:dk-smalldp}).
As commented in subsection~\ref{sec:<k>}, the momentum distribution $\tilde{P}(\tk)$ has a long tail in this parameter regime, and thus $\Delta k$ is exponentially enhanced with the factor $e^{\frac{a^2}{4\Delta p^2}}$ in addition to the classical blue-shift factor $e^{a u_0}$.
This enhancement factor $e^{\frac{a^2}{4\Delta p^2}}$ of $\Delta k$ is larger than that for $\braket{k}$ which is $e^{\frac{a^2}{8\Delta p^2}}$ in \eqref{gaussk-app}. 
Hence, we have $\Delta k \gg \braket{k}$ for $\Delta p \ll a \ll p_0$, and the wave packet cannot have a well-localized momentum.\footnote{$\Delta k \gg \braket{k}$ does not imply that the wave packet contains negative $k$ modes because we assume that the wave packet has only positive $p$ modes and the Bogoliubov transformation \eqref{Bogo-a}, \eqref{Bogo-b} does not mix positive and negative modes. As in Fig.~\ref{fig:p0-200_dp-1}, the momentum distribution $P(k)$ has a long tail for large $k$.}
In particular, a wave packet that is almost a plane wave in the Rindler frame acquires a large momentum fluctuation in the Minkowski frame.

Fig.~\ref{fig:dp_vs_dk} implies that there is a minimum value of the Minkowski momentum uncertainty. 
From \eqref{eq:Deltak-app2}, $\Delta k$ takes the smallest value 
\begin{align}
    \Delta k \simeq e^{a u_0} \sqrt{a p_0}
    \label{eq:del-k}
\end{align}
when $\Delta p \simeq \sqrt{a p_0/2}$ for fixed $a, u_0, p_0$.
No matter how much the Rindler observer reduces $\Delta p$, one cannot prepare a wave packet whose Minkowski momentum uncertainty is arbitrarily small. 
When a Rindler observer with acceleration parameter $a$ emits a massless wave packet with average momentum $p_0$, 
we should choose $\Delta p$ such that $\Delta p/ p_0 \simeq \sqrt{a/(2 p_0)}$ to realize the minimum $\Delta k$.
For a typical value of $a$, e.g., $a \sim 10^{-4}$[eV], the condition is satisfied for example, ($p_0 \sim 10^4$[eV], $\Delta p \sim 1$[eV]) or ($p_0 \sim 10^8$[eV], $\Delta p \sim 10^2$[eV]). 
For a black hole of roughly one solar mass, the parameter $a$ is a tiny value, on the order of $10^{-11}$[eV]. For this small $a$, if we take, for instance, $p_0 \sim 1$[eV], the condition $\Delta p \simeq \sqrt{a p_0/2}$ is satisfied for $\Delta p \sim 10^{-6}$[eV].

%%%%%%%%%%%%%%%%%%%%%%%%%%%%%%%%%%%%%%%%%%%%%%%%%%%%%%%%%%%%%%
%%%%%%%%%%%%%%%%%%%%%%%%%%%%%%%%%%%%%%%%%%%%%%%%%%%%%%%%%%%%%%
\subsection{Uncertainty relation in Minkowski spacetime}
\label{sec:uncert}
%%%%%%%%%%%%%%%%%%%%%%%%%%%%%%%%%%%%%%%%%%%%%%%%%%%%%%%%%%%%%%
%%%%%%%%%%%%%%%%%%%%%%%%%%%%%%%%%%%%%%%%%%%%%%%%%%%%%%%%%%%%%%
We have evaluated the variance $\Delta k$ of the Minkowski momentum $k$. The spatial width $\Delta U$ of the wave packet is also given as \eqref{DeltaU-Deltap-1} and \eqref{DeltaU-Deltap-2}.
Combining these results, we now summarize the resulting position–momentum uncertainty relation.

\paragraph{Case (i): $a \ll \sqrt{a p_0/2} <\Delta p \ll p_0$.}
In this parameter regime, using $\Delta u \simeq 1/\Delta p$, we obtain
\begin{align}
    \Delta U \simeq \frac{1}{2}e^{-a u_0}(\Delta p)^{-1}, \quad \Delta k \simeq e^{a u_0}\Delta p.
\end{align}
Consequently, we have 
\begin{align}
    \Delta U \, \Delta k \simeq \frac{1}{2}.
\end{align}
Therefore, the uncertainty relation $\Delta U \, \Delta k \geq 1/2$ is saturated, independent of $u_0$ (or the location of the wave packet). 
This indicates that the wave packet retains a Gaussian form even in Minkowski coordinates as seen in Fig.~\ref{fig:p0-200_a-1}. 
We further find
\begin{align}
    \frac{\Delta U}{|U_0|}\simeq \frac{a}{2\Delta p} \ll 1, \quad 
    \frac{\Delta k}{\braket{k}} \simeq \frac{\Delta p}{p_0} \ll 1.
\end{align}
where $U_0=-a^{-1}e^{-a u_0}$.
Thus, the wave packet is well localized in the momentum space, and the spatial width $\Delta U$ is much shorter than $|U_0|$. 

If our free-field description is applicable only below a cutoff energy scale $\Lambda$, we should impose $\braket{k} < \Lambda$. 
In the current parameter regime, it leads to
\begin{align}
    1/\Lambda < 1/\braket{k} \ll \Delta U \ll |U_0|.
\end{align}
It indicates that our effective field theory description breaks down unless $ 1/\Lambda \ll |U_0|$.
$\braket{k} < \Lambda$ requires $e^{a u_0}p_0 <\Lambda$; we thus need
\begin{align}
\label{u0:cond1}
    u_0 < a^{-1}\log \frac{\Lambda}{p_0} \ll a^{-1}\log \frac{\Lambda_\text{Pl}}{a}\,,
\end{align}
where $\Lambda_\text{Pl}$ denotes the Planck energy scale and we have supposed $a \ll p_0 <\Lambda <\Lambda_\text{Pl}$. 
Note that $a^{-1}\log (\Lambda_\text{Pl}/a)$ is of the same order as the scrambling time \cite{Sekino:2008he} for a black hole with mass $M\sim a$. 
Hence, our EFT description breaks down before the scrambling time, necessitating an UV-complete description.

If the condition \eqref{u0:cond1} is satisfied, the gravitational back-reaction by this wave packet is negligible as follows. 
The typical Schwarzschild radius scale of this wave packet is $\braket{k} /\Lambda_\text{Pl}^2$, where $\Lambda_\text{Pl}$ is the Planck energy scale.
To avoid gravitational collapse, the spatial width $\Delta U$ must be greater than this radius.
Indeed, we have
\begin{align}
    \frac{\Delta U}{\braket{k} /\Lambda_\text{Pl}^2} \simeq \frac{\Lambda_\text{Pl}^2 e^{-2 a u_0}}{p_0 \Delta p}>\frac{\Lambda_\text{Pl}^2\, p_0}{\Lambda^2 \,\Delta p} \gg 1,
\end{align}
where we have used \eqref{u0:cond1} and supposed $\Lambda <\Lambda_\text{Pl}$.

\paragraph{Case (ii): $a \ll  \Delta p < \sqrt{a p_0/2} \ll p_0$.}
In this parameter regime, we have
\begin{align}
    \Delta U \simeq \frac{1}{2}e^{-a u_0}(\Delta p)^{-1}, \quad \Delta k \simeq e^{a u_0}\frac{a p_0}{2\Delta p}.
\end{align}
Consequently, we have 
\begin{align}
    \Delta U \, \Delta k \simeq \frac{a p_0}{4 (\Delta p)^2}>\frac{1}{2}.
\end{align}
This indicates that the wave packet deviates from a Gaussian form in Minkowski coordinates because the uncertainty relation $\Delta U \, \Delta k \geq 1/2$ is not saturated.
Nevertheless, we have
\begin{align}
    \frac{\Delta U}{|U_0|}\simeq \frac{\Delta k}{\braket{k}} \simeq \frac{a}{2\Delta p} \ll 1,
    \label{U-k_1-2}
\end{align}
and thus the wave packet remains well localized in the momentum space, independent of $u_0$.
Note that the ratios $\Delta U/|U_0|$, $\Delta k/\braket{k}$ are of the same order of magnitude and are independent of $p_0$.

Imposing $\braket{k} < \Lambda$, we have
\begin{align}
    1/\Lambda < 1/\braket{k}  \ll \Delta U \ll |U_0|
\end{align}
as in the case (i), and we must require 
\begin{align}
    u_0 < a^{-1}\log \frac{\Lambda}{p_0} \ll a^{-1}\log \frac{\Lambda_\text{Pl}}{a}\,.
\end{align}

This parameter regime is also noteworthy, as mentioned in section~\ref{sec:Var-k},
because the uncertainty relation $\Delta k\, \Delta p \geq  \frac{a}{2}\braket{k}$ is saturated.
Since \eqref{U-k_1-2} leads to
\begin{align}
    \Delta U \simeq \frac{|U_0|}{\braket{k}} \Delta k,
\end{align}
the smaller $\Delta k$ is, the smaller $\Delta U$ becomes. 
The minimum value of $\Delta k$ is given by $\Delta k \sim e^{a u_0} \sqrt{a p_0}$ as shown in \eqref{eq:del-k} for fixed $a, u_0, p_0$.
Thus, the minimum value of $\Delta U$ in this parameter regime is $\Delta U \simeq  e^{-a u_0}/\sqrt{2 a p_0}$.

\paragraph{Case (iii): $\Delta p \ll a  \ll p_0$.}
In this parameter regime, we have
\begin{align}
    \Delta U \simeq 
    e^{-a u_0+\frac{a}{4\Delta p}}  a^{-1}
    , \quad \Delta k \simeq e^{a u_0+\frac{a^2}{4\Delta p^2}}p_0.
\end{align}
Consequently, we have 
\begin{align}
    \Delta U \, \Delta k \simeq e^{\frac{a^2}{4\Delta p^2}+\frac{a}{4\Delta p}}\,\frac{p_0}{a} \gg \frac{1}{2}.
\end{align}
Thus, the wave packet differs significantly from a Gaussian profile in Minkowski coordinates.
Indeed, we have
\begin{align}
    \frac{\Delta U}{|U_0|} \simeq e^{\frac{a}{4\Delta p}} \gg 1, \quad  \frac{\Delta k}{\braket{k}} \simeq e^{\frac{a^2}{8\Delta p^2}} \gg 1.
\end{align}
The wave packet is not localized in the Minkowski momentum space as mentioned in sec.~\ref{sec:Var-k}.
The spatial wave profile is also distorted such as it appears to be stuck to the horizon. 
Under the coordinate transformation $U=-a^{-1}e^{-a u}$, the right part ($u >u_0$) of the wave function in Rindler coordinates is compressed in Minkowski coordinates, while the left part ($u <u_0$) is extended in Minkowski coordinates.
Hence, $U_0=-a^{-1}e^{-a u_0}$  does not represent the average value of $U$, although $U_0$ still represents a position of the wave-packet peak. 

Imposing $\braket{k} < \Lambda$, we have
\begin{align}
    1/\Lambda < 1/\braket{k}  \ll |U_0| \ll \Delta U .
\end{align}
The validity of EFT requires $\braket{k} +\Delta k< \Lambda$, and it leads to $e^{a u_0+\frac{a^2}{4\Delta p^2}}p_0 \lesssim \Lambda$, that is, 
\begin{align}
    u_0 < -\frac{1}{a}\left[\frac{a^2}{4\Delta p^2}-\log \frac{\Lambda}{p_0}\right]\,.
\end{align}
Since the second term $\log (\Lambda/p_0)$ is logarithmic, it may be natural to think that the magnitude is not so large compared to the first term.\footnote{For example, for $p_0 \sim 1$[eV], $a \sim 10^{-4}$[eV] and $\Delta p \sim 10^{-6}$[eV], we have $a^2/(4\Delta p^2) \sim \mathcal{O}(10^3) \gg \log (\Lambda_\text{Pl}/p_0) \sim \mathcal{O}(10)$. 
Thus, even if we take $\Lambda$ as the Planck scale, the second term is negligible.
Then, we encounter the breakdown of EFT for $u_0 \gtrsim-10^{-3}$[m].} 
It indicates that the EFT description breaks down even in regions away from the horizon.

%%%%%%%%%%%%%%%%%%%%%%%%%%%%%%%%%%%%%%%%%%%%%%%%%%%%%%%%%%%%%%
%%%%%%%%%%%%%%%%%%%%%%%%%%%%%%%%%%%%%%%%%%%%%%%%%%%%%%%%%%%%%%
\section{Conclusion}\label{sec:concl}
%%%%%%%%%%%%%%%%%%%%%%%%%%%%%%%%%%%%%%%%%%%%%%%%%%%%%%%%%%%%%%
%%%%%%%%%%%%%%%%%%%%%%%%%%%%%%%%%%%%%%%%%%%%%%%%%%%%%%%%%%%%%%
We have studied (Gaussian) wave packets with localized Rindler momenta constructed on the Minkowski vacuum, and have examined how they appear to Minkowski observers. 
Wave packets with localized Rindler momenta should have $p_0 \gg \Delta p, a$.
We have considered this parameter regime and seen whether the classical blue-shift formula $k=e^{a u_0} p_0$ is corrected.
For wave packets with $a \ll \Delta p \ll p_0$, the expectation value of the Minkowski momentum reproduces the classical result $\braket{k}\simeq e^{a u_0}p_0$. 
More precisely, for $a \ll \sqrt{a p_0/2} <\Delta p \ll p_0$, the wave packets appear to be the Gaussian wave packets in the Minkowski spacetime, while for $a \ll \Delta p<\sqrt{a p_0/2}  \ll p_0$ they do not.
Nevertheless, in the latter parameter regime, the wave packets almost saturate the uncertainty relation between the Minkowski and Rindler momentum $\Delta k\, \Delta p  \geq \frac{a}{2}\braket{k}$.
For $\Delta p \lesssim a$, the blue shift is enhanced as
$\braket{k} \simeq e^{a u_0+\frac{a^2}{8\Delta p^2}}p_0$ compared to the classical result, worsening the trans-Planckian problem, although the wave packets differ significantly from Gaussian profiles in the Minkowski frame.

We have also found that the variance of the Minkowski momentum $\Delta k$ has a lower bound $ \Delta k \gtrsim e^{a u_0} \sqrt{a p_0}$, no matter how small $\Delta p$ is, although one may think that we can make $\Delta k$ arbitrarily small by reducing $\Delta p$.
The bound is achieved when $\Delta p \simeq \sqrt{a p_0/2}$. 

For future work, it would be worthwhile to evaluate the back-reaction of the Rindler wave-packet states by computing the expectation value of the stress-energy tensor $\bra{\psi_R} T_{\mu\nu}(x)\ket{\psi_R}$. 
The blue-shift of energy may cause a large back-reaction, though we are not sure whether the EFT description is valid in this situation.
It would also be illuminating to compare the back-reacted geometry to the shock wave geometry \cite{Dray:1984ha, Dray:1985yt} for various parameter regimes.
The shock wave geometry is also related to quantum chaos properties of black holes \cite{Shenker:2013pqa}. 
Thus, it would also be valuable to compute OTOC-like correlators $\bra{\psi_R} \phi(x) \phi(x')\ket{\psi_R}$.

Related to this, a promising next step is to compute the black hole $S$-matrix \cite{tHooft:1987vrq, tHooft:1996rdg} using wave packets. 
In \cite{tHooft:1996rdg}, it is asserted that the position of shock waves and the horizon satisfy an uncertainty relation.
The uncertainty might be related to the above lower bound of $\Delta k$.
Adding the gravitational interaction, the wave-packet states must involve the gravitational dressing \cite{Donnelly:2015hta, Donnelly:2016rvo, Sugishita:2024lee} to satisfy the Gauss law constraint for gravity. 
The dressing should be related to the asymptotic symmetry \cite{Bondi:1962px, Sachs:1962wk, Sachs:1962zza}, and thus it is related to the gravitational memory effect \cite{Zeldovich:1974gvh, Braginsky:1985vlg, Braginsky:1987kwo}, as in QED (see for instance \cite{Hirai:2022yqw}).\footnote{Furthermore, the dressing to wave packets is inevitable to obtain the non-zero amplitude due to the superselection rule of asymptotic symmetry as argued in \cite{Carney:2018ygh, Hirai:2022yqw}.}
Indeed, it is argued in  \cite{He:2023qha} that the memory effect is related to ’t Hooft’s uncertainty relation.
In this sense, the evaluation of $\bra{\psi_R} T_{\mu\nu}(x)\ket{\psi_R}$ is also important.

The trans-Planckian problem suggests that UV physics is important near the horizon. 
Extending our analysis beyond the free QFT is worth pursuing, for example, by adding the higher derivative interaction as \cite{Ho:2021sbi, Ho:2022gpg}.
Even in free QFT, it is an appealing avenue to generalize our analysis to the case with an IR cutoff like the brick wall model \cite{tHooft:1984kcu}, or the case with a minimal length or an UV cutoff \cite{Garay:1994en, Kempf:1994su, Mu:2015qta, Ho:2022gpg, Chau:2023zxb}.
In particular, if we introduce an UV cutoff to the Minkowski momentum, it is not clear how the Rindler momentum is restricted. 
The restriction leads to the change of the Bogoliubov transformation formula \eqref{eq:bogo-ba}, and the Unruh thermal state may be modified. 

The brick wall model \cite{tHooft:1984kcu} or the fuzzball proposal \cite{Mathur:2005zp} implies that the semi-classical spacetime picture is invalid near the horizon. 
Based on \cite{Sugishita:2023wjm}, the situation is drastically different depending on whether it is a single-sided black hole corresponding to the gravitational collapse or the two-sided eternal black holes. 
Our Minkowski analysis corresponds to the two-sided case. Thus, investigating a single-sided case would be of considerable interest.

Finally, it is highly motivating to consider observable predictions and phenomenological applications of quantum gravity by using the gravitational blue-shift. 
By utilizing red/blue-shift effects, one may be able to observe high-energy physics.
Extremely microscopic physics near the horizon may become much longer wavelengths and time scales when viewed from far away.
It may offer a window into quantum gravitational effects, allowing them to become accessible to experiments in the foreseeable future.

%%%%%%%%%%%%%%%%%%%%%%%%%%%%%%%%%%%%%%%%%%%%%%%%%%%%%%%%%%%%%%
%%%%%%%%%%%%%%%%%%%%%%%%%%%%%%%%%%%%%%%%%%%%%%%%%%%%%%%%%%%%%%
\section*{Acknowledgement}
%%%%%%%%%%%%%%%%%%%%%%%%%%%%%%%%%%%%%%%%%%%%%%%%%%%%%%%%%%%%%%
%%%%%%%%%%%%%%%%%%%%%%%%%%%%%%%%%%%%%%%%%%%%%%%%%%%%%%%%%%%%%%
Sotaro Sugishita acknowledges support from JSPS KAKENHI Grant Numbers JP21K13927 and JP22H05115, and JST BOOST Program Japan Grant Number JPMJBY24E0.

%%%%%%%%%%%%%%%%%%%%%%%%%%%%%%%%%%%%%%%%%%%%%%%%%%%%%%%%%%%%%%
%%%%%%%%%%%%%%%%%%%%%%%%%%%%%%%%%%%%%%%%%%%%%%%%%%%%%%%%%%%%%%
\appendix
%%%%%%%%%%%%%%%%%%%%%%%%%%%%%%%%%%%%%%%%%%%%%%%%%%%%%%%%%%%%%%
\section{Rindler space as near-horizon geometry of black holes}
\label{app:Rind}
%%%%%%%%%%%%%%%%%%%%%%%%%%%%%%%%%%%%%%%%%%%%%%%%%%%%%%%%%%%%%%
Here we review that the Rindler spacetime is obtained by taking the near-horizon limit of the Schwarzschild black hole geometry.

The metric of the four-dimensional Schwarzschild black hole is
\begin{align}
    ds^2=-\left(1-\frac{r_s}{r}\right)dt^2+\left(1-\frac{r_s}{r}\right)^{-1}dr^2+r^2 d\Omega^2,
\end{align}
where $r_s$ is the Schwarzschild radius: $r_s=2 G_N M$ for a black hole with mass $M$. 
We consider the near-horizon region $r = r_s +\epsilon$ with $\epsilon \ll r_s$.
Setting $\epsilon = r_s e^{x/r_s}$, we find that the metric is approximated as
\begin{align}
    ds^2= e^{x/r_s}(-dt^2+dx^2)+r_s^2 d\Omega^2.
\end{align}
The spherical part $r_s^2 d\Omega^2$ can be negligible if we consider only the s-wave part of massless fields. 
The other two-dimensional part $e^{x/r_s}(-dt^2+dx^2)$ is the two-dimensional Rindler spacetime with $a=1/(2 r_s)=1/(4 G_N M)$, which is the surface gravity. 

To ensure the near-horizon condition $\epsilon = r_s e^{x/r_s} \ll r_s$, we need $x<0$ and $|x| \gg r_s$.
For fixed $u$, it means $t \ll u-r_s$, and for fixed $v$, it means $t\gg v+r_s$.

%%%%%%%%%%%%%%%%%%%%%%%%%%%%%%%%%%%%%%%%%%%%%%%%%%%%%%%%%%%%%%
\section{Bogoliubov transformation}
\label{app:Bogo}
%%%%%%%%%%%%%%%%%%%%%%%%%%%%%%%%%%%%%%%%%%%%%%%%%%%%%%%%%%%%%%
Here we review the Bogoliubov transformation, which we use in this paper, between the Minkowski and Rindler modes for the free massless scalar field in two dimensions based on \cite{Crispino:2007eb}.

Both of the modes $\left\{e^{\pm i(\omega_k T-k X)}\right\}$ in \eqref{phi-Mink} and $\left\{e^{\pm i(\omega_p t-p x)}\right\}$ in \eqref{phi-Rind} are complete on the right Rindler wedge, and thus the Rindler modes $\left\{e^{\pm i(\omega_p t-p x)}\right\}$ can be expanded by the Minkowski modes $\left\{e^{\pm i(\omega_k T-k X)}\right\}$ via the Bogoliubov transformation as follows. 
\begin{align}
    \frac{1}{\sqrt{4\pi\omega_p}} e^{-i(\omega_p t-p x)}&=\int^\infty_{-\infty}\frac{dk}{\sqrt{4\pi\omega_k}}\left[ \alpha_{p,k} e^{-i(\omega_k T-k X)}+ \beta_{p,k}e^{+i(\omega_k T-k X)}\right],
    \label{eq:bogo-}\\
    \frac{1}{\sqrt{4\pi\omega_p}} e^{+i(\omega_p t-p x)}&=\int^\infty_{-\infty}\frac{dk}{\sqrt{4\pi\omega_k}}\left[ \beta_{p,k}^\ast  e^{-i(\omega_k T-k X)}+ \alpha_{p,k}^\ast e^{+i(\omega_k T-k X)}\right].
\end{align}
From the above relation, one finds the transformation rule between $a_k$ and $b_p$ as
\begin{align}
    b_{p}=\int^{\infty}_{-\infty} dk \left[ \alpha_{p,k}^\ast a_k- \beta^\ast_{p,k} a_k^\dagger\right], \qquad b^\dagger_{p}=\int^{\infty}_{-\infty} dk \left[ \alpha_{p,k} a_k^\dagger- \beta_{p,k} a_k\right].
    \label{eq:bogo-ba}
\end{align}

One can explicitly obtain the Bogoliubov coefficients \cite{Crispino:2007eb},
\begin{align}
    \alpha_{p,k}&=[\theta(p)\theta(k)+\theta(-p)\theta(-k)]
    \frac{e^{\frac{\pi \omega_p}{2a}}}{2\pi a}\sqrt{\frac{\op}{\ok}}
    \left(\frac{a}{\ok}\right)^{\frac{i\op}{a}}\Gamma(i\op/a),
    \label{Bogo-a}
    \\
    \beta_{p,k}&=[\theta(p)\theta(k)+\theta(-p)\theta(-k)]
    \frac{-e^{-\frac{\pi \omega_p}{2a}}}{2\pi a}\sqrt{\frac{\op}{\ok}}
    \left(\frac{a}{\ok}\right)^{\frac{i\op}{a}}\Gamma(i\op/a).
    \label{Bogo-b}
\end{align}
In our paper, we use the form of the coefficients only for $p>0, k>0$, and they read
\begin{align}
    \alpha_{p,k}=
    \frac{e^{\frac{\pi p}{2a}}}{2\pi a}\sqrt{\frac{p}{k}}
    \left(\frac{a}{k}\right)^{\frac{i p}{a}}\Gamma(i p/a),\quad
    \beta_{p,k}=
    \frac{-e^{-\frac{\pi p}{2a}}}{2\pi a}\sqrt{\frac{p}{k}}
    \left(\frac{a}{k}\right)^{\frac{i p}{a}}\Gamma(i p/a).
\end{align}
The absolute values are given by
\begin{align}
      |\alpha_{p,k}|^2   =\frac{e^{\frac{2\pi p}{a}}}{2\pi a \omega_k(e^{\frac{2\pi p}{a}}-1)},\quad 
      |\beta_{p,k}|^2   =\frac{1}{2\pi a \omega_k(e^{\frac{2\pi p}{a}}-1)}
\end{align}
for $p>0, k>0$. 
For the Minkowski vacuum $\ket{0}_M$ (satisfying $a_k \ket{0}_M=0$), we have
\begin{align}
    \tensor[_M]{\bra{0}}{}b^\dagger_p b_{p'}\ket{0}_M=&\int^\infty_0dk \beta_{p,k}\beta^\ast_{p',k}=\frac{1}{e^{\frac{2\pi p}{a}}-1}\delta(p-p'),
        \\
    \tensor[_M]{\bra{0}}{}b_{p'} b^\dagger_{p}\ket{0}_M=&\int^\infty_0dk \alpha^\ast_{p',k}\alpha_{p,k}=\frac{e^{\frac{2\pi p}{a}}}{e^{\frac{2\pi p}{a}}-1}\delta(p-p'),
    \label{intk-aa}
\end{align}
where we have supposed $p>0, p'>0$. 
% We denote the Planck distribution by $B(p)$,
% \begin{align}
%     B(p):=\frac{1}{e^{\frac{2\pi p}{a}}-1}.
% \end{align}

%%%%%%%%%%%%%%%%%%%%%%%%%%%%%%%%%%%%%%%%%%%%%%%%%%%%%%%%%%%%%%
\section{Saddle point approximation of the momentum distribution}
\label{app:LambertW}
%%%%%%%%%%%%%%%%%%%%%%%%%%%%%%%%%%%%%%%%%%%%%%%%%%%%%%%%%%%%%%
%%%%%%%%%%%%%%%%%%%%%%%%%%%%%%%%%%%%%%%%%%%%%%%%%%%%%%%%%%%%%%
In this appendix, we evaluate $I_+(\tk)$ defined by \eqref{def:I+} for the Gaussian-wave packets, i.e., for the case that $f_{p_0,\Delta p}$ in \eqref{def:I+} is given by
\begin{align}
    f_{p_0,\Delta p}(p)=\frac{1}{(2\pi \Delta p^2)^{\frac{1}{4}}}e^{-\frac{1}{4}\left(\frac{p-p_0}{\Delta p}\right)^2}.
\end{align}

As explained in subsection~\ref{subsec:wp-Rindler}, we are interested in the case that $p_0 \gg a$. 
We also suppose that $\Delta p$ is less than $p_0$, i.e., we suppose $p_0 \gg \Delta p$ because we are considering wave packets such that they approximately have a definite momentum $p_0$. 
Then, in the $p$-integral 
\begin{align}
I_+(\tk):=\sqrt{a}\int_{0}^{\infty}dp\,f_{p_0,\Delta p}(p)  \frac{e^{\frac{\pi p}{2a}}}{2\pi a}\tk^{-\frac{i p}{a}}
\sqrt{\frac{p}{a}}\,
   \Gamma\!\left(\frac{i p}{a}\right),
\end{align}
we can regard $p$ is much greater than $a$ because $f_{p_0,\Delta p}$ is almost supported in the region $p \gtrsim p_0-\Delta p \gg a$.
We can thus use the approximation
\begin{align}
    \sqrt{\frac{p}{a}}\,
   \Gamma\!\left(\frac{i p}{a}\right)\simeq \sqrt{2\pi}e^{ i\frac{p}{a}\log (p/a)-\frac{\pi p}{2a}-\frac{ip}{a}-\frac{i\pi}{4}}
\end{align}
for $p \gg a$, and obtain
\begin{align}
I_+(\tk)&\simeq\frac{e^{-\frac{i\pi}{4}}}{(2\pi)^{3/4}}\left(\frac{a}{\Delta p}\right)^{\frac{1}{2}}\int_{0}^{\infty}\frac{dp}{a}\,e^{g(p)},
\label{Ik+eg}
\\
g(p)&:=-\frac{1}{4}\left(\frac{p-p_0}{\Delta p}\right)^2
+i\frac{p}{a}\left[\log (p/a \tk)-1\right].
\end{align}
We now use the saddle point approximation.
Let $p_\ast$ be the solution of $g'(p)=0$, i.e.,
\begin{align}
    -\frac{p_\ast-p_0}{2 \Delta p^2}
+\frac{i}{a}\log (p_\ast/a \tk)=0,
\end{align}
which means 
\begin{align}
   \frac{i a p_\ast}{2\Delta p^2}e^{\frac{i a p_\ast}{2\Delta p^2}}
   =\frac{i a^2 \tk}{2\Delta p^2}e^{\frac{i a p_0}{2\Delta p^2}}.
\end{align}
Thus, $p_\ast$ can be expressed by
the Lambert $W$ function, i.e. the inverse function of $z=w e^w$, which is expressed as $w=W_n(z)$ where $n$ takes integers representing the branches.
We now have
\begin{align}
    p_\ast=-\frac{2i \Delta p^2}{a}W_{n}\left(\frac{i a^2 \tk}{2\Delta p^2}e^{\frac{i a p_0}{2\Delta p^2}}\right),
\end{align}
where
$n$ is chosen so that $p_\ast =a \tk e^{\frac{i a p_0}{2\Delta p^2}}$ in the limit $a/\Delta p \to 0$ for finite $\tk$ and $\frac{a p_0}{\Delta p^2}<\pi$, 
and the appropriate $n$ depends on $p_0, \Delta p, a$.
To be concrete, $n$ is given by
\begin{align}
    n=\left\lceil \frac{\frac{a p_0}{2\Delta p^2}-\frac{\pi}{2}}{2\pi} \right\rceil,
\end{align}
where $\lceil \cdot \rceil$ is the ceiling function: $\lceil x\rceil:=\min\left\{n \in \mathbb{Z}\,|\, x \leq n\right\}$.

We then have
\begin{align}
    g(p_\ast)=\frac{p_\ast^2-p_0^2}{4\Delta p^2}-\frac{ip_\ast}{a}, \qquad     
    g''(p_\ast)=-\frac{1}{2 \Delta p^2}+\frac{i}{a p_\ast}. 
\end{align}
The saddle point approximation of \eqref{Ik+eg} leads to
\begin{align}
|I_+(\tk)|&\simeq |I^\text{saddle}_+(\tk)|:=\frac{e^{\mathrm{Re}[g(p_\ast)]}}{(2\pi)^{1/4}\left|\frac{a}{2\Delta p}-\frac{i \Delta p}{p_\ast}\right|^{\frac{1}{2}}},
\end{align}
where we take the absolute value because we need only $|I_+(\tk)|$ for the evaluation of $\tilde{P}(\tk)$ in \eqref{def:tilP}.
%%%%%%%%%%%%%%%%%%%%%%%%%%%%%%%%%%%%%%%%%%%%%%%%%%%%%%%%%%%%%%

%%%%%%%%%%%%%%%%%%%%%%%%%%%%%%%%%%%%%%%%%%%%%%%%%%%%%%%%%%%%%%
%%%%%%%%%%%%%%%%%%%%%%%%%%%%%%%%%%%%%%%%%%%%%%%%%%%%%%%%%%%%%%
\bibliographystyle{utphys}
\bibliography{ref-wave}
%%%%%%%%%%%%%%%%%%%%%%%%%%%%%%%%%%%%%%%%%%%%%%%%%%%%%%%%%%%%%%
\end{document}